\documentclass[reprint,preprintnumbers, amsmath,amssymb, aps]{revtex4-2}

\usepackage{graphicx}
\usepackage{dcolumn}
\usepackage{bm}
\usepackage{changes}
\usepackage{soul}
\usepackage{amsmath}
\usepackage{color}

\DeclareUnicodeCharacter{3000}{\space}

\begin{document}

\title{Scattering features of a self-dual spectral singularity in coupled ring resonators with balanced gain and loss}

\author{Jeng Yi Lee }

\affiliation{Department of Opto-Electronic Engineering, National Dong Hwa University, Hualien 974301, Taiwan
}

\begin{abstract}
We theoretically and numerically discuss the general scattering property of a coupled two-ring-resonators externally coupled to a straight bus waveguide.
Specifically, we require one ring resonator to be light field amplifying and another to be that dissipative, where the associated attenuation and gain factors for a field amplitude after one-round trip are formed in a reciprocal pair as well as the effective wave propagation constants  are same.
Obtained from reciprocity and an effective PT-symmetric transfer matrix, we exploit parametrization to propose a general phase diagram, identifying  overall scattering outcomes.
Especially, we find that the formation of  a self-dual spectral singularity, namely coherent perfect absorption-lasing (CPAL), is occurred at a certain relation of a coupling coefficient and an attenuation factor as well as  resonant whispering gallery  modes.
Reducing light amplification and dissipation strengths in each rings narrows the spectral bandwidth associated with CPAL.
We further discover that any system with the CPAL parameters would exhibit a same scattering feature: transmission extinction, CPAL, and uni-directional reflectionless at the loss ring side, with respect to an incident wavelength.
Additionally, a transmission extinction can be away from the resonance.
These analytical predictions are verified by electromagnetic simulations. 
\end{abstract}

\maketitle

 \section{Introduction}
 Optical ring resonators are key components for on-chip photonic integrated circuits, with diverse applications across science and engineering, including optical memory \cite{memory1}, sensing \cite{Wiersig2011,Chen2017}, quantum controlled-NOT gates \cite{Scott2019},  optical squeezing \cite{Dutt2015}, and wavelength division multiplexing \cite{wavelength1} to name a few examples.
 
A ring resonator externally coupled to a straight 
bus waveguide exhibits the critical coupling at a whispering-gallery resonance when an intrinsic loss balances an external coupling, as demonstrated experimentally \cite{critical1}. 
Destructive interference between a directly transmitted wave and  a circulating wave in a ring externally coupling to a bus waveguide results in a transmission extinction. 
This scattering event with an absence of both reflection and transmission uniquely represents a single-sided coherent perfect absorption, corresponding to a zero eigenvalue of the scattering matrix \cite{cpa1}.
In a two-port slab system, coherent perfect absorption
generally requires two counterpropagating incident waves with a prescribed relative phase and amplitude
ratio~\cite{Baranov2017,Wan2011}.
 
Optical gain in ring resonators can be provided by
electrically pumped quantum dots~\cite{Wan2017}
or multiple quantum wells~\cite{Liang2009},
optically pumped rare-earth ions~\cite{Luo2021},
or nonlinear processes such as stimulated Raman
scattering~\cite{Rong2007} and Kerr four-wave-mixing-based
parametric amplification~\cite{Levy2010}.
Moreover, a lasing threshold in a single active ring system is reached when an amplification factor for a field amplitude after one round-trip equals a reciprocal of a through coupling coefficient at resonance, corresponding to a pole eigenvalue of the associated scattering matrix.
Electrically pumped microring lasing has been
experimentally demonstrated on a hybrid III--V/silicon platform~\cite{Liang2009}. 

  \begin{figure*}[ht]
 \centering
\includegraphics[width=0.8\textwidth]{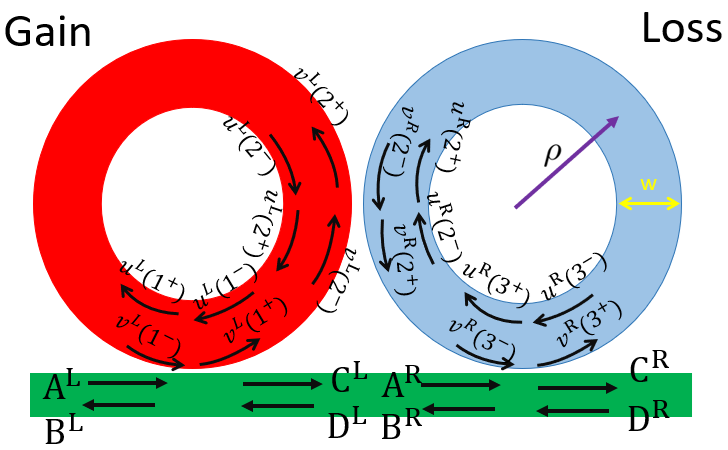}
\caption{Schematic of two mutually coupled ring resonators,
with gain in the left ring ($L$) and loss in the
right ring ($R$), each coupled to a common straight
bus waveguide.
Both rings have radius $\rho$ and waveguide width $w$.
The clockwise and counterclockwise circulating field
amplitudes are denoted by $u^{L,R}$ and $v^{L,R}$,
respectively.
Coupling regions $1$, $2$, and $3$ correspond to the
left ring--bus, inter-ring, and right ring--bus
couplers, respectively.
The labels $-$ and $+$ indicate the input and output
sides of each coupler for the corresponding propagation
direction.
In the two-port bus waveguide, $A^{L,R}$ and $C^{L,R}$
denote right-propagating field amplitudes, whereas
$B^{L,R}$ and $D^{L,R}$ denote left-propagating
field amplitudes. }
\label{fig1}
\end{figure*}

At discrete real frequencies, spectral singularities
correspond to lasing thresholds, whereas their
time-reversed counterparts correspond to coherent
perfect absorption~\cite{Mostafazadeh2012PRA}.
When both occur at the same frequency, the system
supports coherent perfect absorption and lasing
(CPAL), associated with a self-dual spectral
singularity~\cite{Mostafazadeh2012SelfDual}.
These scattering results belong to non-Hermitian, where relies on material gain and loss embedded.
In particularly, one well-known example with a refractive-index distribution satisfying $n(\vec{r})=n^{*}(-\vec{r})$ signs a parity--time ($\mathcal{PT}$) symmetry, with an even real part and an odd imaginary part corresponding to spatially balanced gain and loss \cite{Guo2009,Makris2008}. 
A single microring resonator with a suitably designed $\mathcal{PT}$-symmetric grating that is externally coupled to a bus waveguide can support simultaneous coherent perfect absorption and lasing (CPAL) at discrete frequencies \cite{Longhi2014}. 
At each CPAL point, a scattering-matrix zero and pole coincide at the same real frequency, forming a self-dual spectral singularity.
 Related studies have demonstrated that due to an unbalanced coupling strength between clockwise and counterclockwise modes of the ring, that  suppresses circulation in one direction while enhancing it in the other \cite{Choi2025}.
In a coupled two ring system, at the $\mathcal{PT}$ symmetry broken phase, it can localize a supermode predominantly in the gain ring, enabling lasing when its net gain exceeds threshold \cite{Konotop2019}, as experimentally demonstrated in an optically pumped InGaAsP quantum-well platform \cite{Hodaei2015,lasing}.
For the aforementioned systems, transmission through
the bus waveguide remains reciprocal for incidence
from either side in both the $\mathcal{PT}$-symmetric
and $\mathcal{PT}$-broken phases, given that the
system operates in the linear regime. 
However, in some broken-symmetry phases, as a result of a strong field localization occurred in the gain ring further enhances nonlinear effects, enabling nonreciprocal transmission \cite{Peng2014}.

In this work, we investigate two mutually coupled ring resonators with one light wave amplifying and  another light wave dissipative, externally coupled to a common straight bus waveguide, as shown in Fig. 1.
We consider identical wave propagation constants in the two rings, an reciprocal pair of  amplification and attenuation factors, and same unitary directional couplers describing evanescent coupling regimes. 
Under these conditions as well as analyzing from a transfer-matrix formulation, we show that the system exhibits effective $\mathcal{PT}$-symmetry even when the refractive-index distribution used in our full-wave electromagnetic simulations does not meet $n(\vec{r})=n^{*}(-\vec{r})$.

Reciprocity in our system is conserved. 
Consequently, we can parameterize the transfer matrix in terms of the coupler through-coupling coefficient, the attenuation factor, and the propagation phase.
As a result, we propose a general phase digram to identify $\mathcal{PT}$ symmetry phase, symmetry breaking point, and $\mathcal{PT}$ symmetry broken phase.
Moreover, we also derive analytical conditions for a transmission extinction, coherent perfect absorption and lasing (CPAL), and unidirectional reflectionlessness. 
Transmission extinction and unidirectional reflectionlessness can occur away from resonance, whereas CPAL occurs at a whispering-gallery
resonance when the coupling coefficient and the
round-trip attenuation factor satisfy a specific relation.
Notably, systems satisfying the CPAL parameter condition
exhibit a common sequence of scattering feature as the
propagation phase varies: transmission extinction,
unidirectional reflection for incidence from the loss-ring
side, and CPAL.
This sequence recurs periodically in the propagation phase
and, when dispersion is neglected in frequency.
We validate these analytical predictions using full-wave electromagnetic simulations in COMSOL Multiphysics.
Our work provides design guidelines for implementing coupled ring resonators with unique scattering integrated into silicon-photonic platforms.

 \begin{figure*}[ht]
 \centering
\includegraphics[width=1\textwidth]{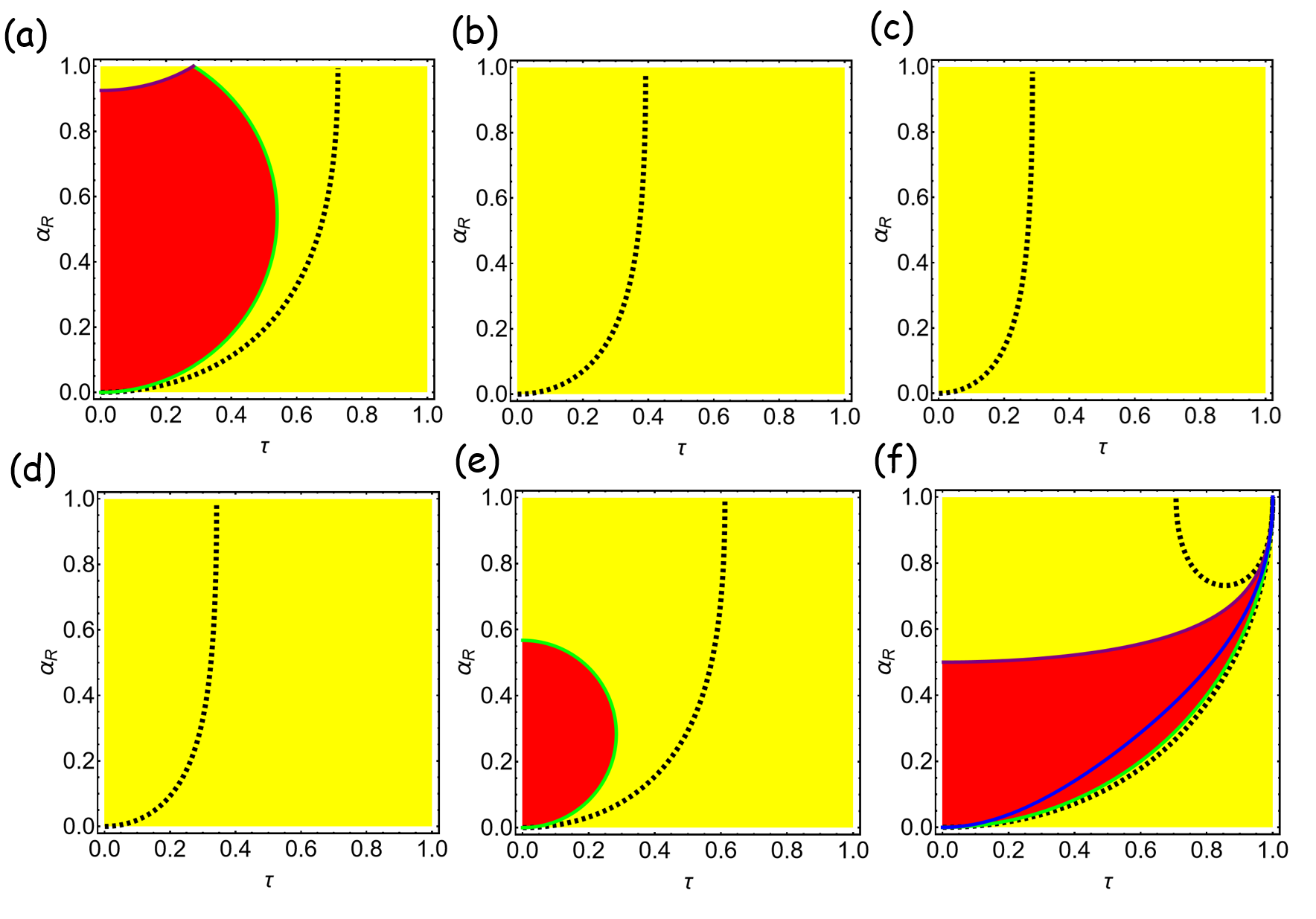}
\caption{Scattering phase diagrams in the $(\tau,\alpha_R)$
plane for selected values of the round-trip propagation
phase $\varphi\equiv 2\pi k_R\rho$:
(a) $\varphi=1$, (b) $\varphi=2$, (c) $\varphi=3$,
(d) $\varphi=4$, (e) $\varphi=5$, and
(f) $\varphi=2\pi$.
Here, $\tau$ is the through-coupling coefficient and
$\alpha_R$ is the round-trip attenuation factor.
Yellow and red regions denote the
$\mathcal{PT}$-symmetric ($T<1$) and
$\mathcal{PT}$-broken ($T>1$) scattering phases,
respectively.
The purple and green curves mark exceptional-point
boundaries ($T=1$).
Black dashed curves indicate transmission extinction
($T=0$), while the blue curve in (f) indicates
coherent perfect absorption and lasing (CPAL). }
\label{fig2}
\end{figure*}

  \section{Theory}
 We consider two mutually coupled ring resonators, each
externally coupled to a common straight bus waveguide,
as shown schematically in Fig.~\ref{fig1}.
Both ring resonators have the same radius $\rho$ and width $w$, with the left ring containing a gain medium and the right ring containing a lossy medium. 
The bus waveguide has two ports, allowing excitation from either side. 
All waveguides are assumed to be operated at a single mode. 
The light field can be approximately represented by $E e^{i(\phi-\omega t)}$, where  $E$ is the complex field
amplitude and $\phi$ is the effective propagation phase.
Throughout this work, we adopt the time dependence $e^{-i\omega t}$.

The system contains three wave coupling regions, labeled $1$, $2$, and $3$, where adjacent waveguides interact
through overlapping evanescent fields.
Each coupling region is modeled as a lossless directional
coupler as well as is described by the same symmetric and unitary
coupling matrix $\mathcal{T}=\left[\begin{smallmatrix}
\tau & i\kappa\\
i\kappa & \tau
\end{smallmatrix}\right]$.
This unitary coupling matrix had been experimentally demonstrated in Ref.~\cite{critical1}.
We note that $\tau$ and $\kappa$ are the real amplitude through- and cross-coupling coefficients, respectively, also satisfying $\tau^2+\kappa^2=1$.

Each ring supports clockwise and counterclockwise circulating modes, denoted by $u^{L,R}$ and $v^{L,R}$, respectively, where the superscripts $L$ and $R$ represent the left and right rings. 
At the inter-ring coupling region, the clockwise mode in either ring couples to the counterclockwise mode in the other, yielding $\left[u^{L}(2^{+}),v^{R}(2^{+})\right]^{T}=\mathcal{T}\left[u^{L}(2^{-}),v^{R}(2^{-})\right]^{T}$ and $\left[v^{L}(2^{+}),u^{R}(2^{+})\right]^{T}=\mathcal{T}\left[v^{L}(2^{-}),u^{R}(2^{-})\right]^{T}$. 
Here, the superscripts $-$ and $+$ on the coupling-region denote the input and output sides, respectively, for each propagation direction.

At the coupling region between the left ring and the bus waveguide,  the clockwise and counterclockwise modes satisfy
$
\left[u^{L}(1^{+}),B^{L}\right]^{T}=\mathcal{T}\left[u^{L}(1^{-}),D^{L}\right]^{T}$ and
$\left[v^{L}(1^{+}),C^{L}\right]^{T}
=\mathcal{T}\left[v^{L}(1^{-}),A^{L}\right]^{T}$, respectively.
Similarly, at the coupling region between the right ring
and the bus waveguide, we have
$\left[u^{R}(3^{+}),B^{R}\right]^{T}
=\mathcal{T}\left[u^{R}(3^{-}),D^{R}\right]^{T}$, and $
\left[v^{R}(3^{+}),C^{R}\right]^{T}
=\mathcal{T}\left[v^{R}(3^{-}),A^{R}\right]^{T}$, respectively.

We denote the real effective propagation constants in the left and right rings by $k_L$ and $k_R$, respectively. 
The corresponding  field amplitude factors, after one round-trip, are $\alpha_L>1$ for the amplifying left ring and $\alpha_R<1$ for the dissipative right ring. 
We impose equal propagation constants $k_L=k_R$ and reciprocal
round-trip amplitude factors $\alpha_L\alpha_R=1$.

Combining these coupling relations and required conditions  yields
the overall transfer matrix relation:  \begin{equation}
\begin{pmatrix}
C^{R}\\
D^{R}
\end{pmatrix}
=M
\begin{pmatrix}
A^{L}\\
B^{L}
\end{pmatrix},
\qquad
M=
\begin{pmatrix}
M_{11} & M_{12}\\
M_{21} & M_{22}
\end{pmatrix}.
\end{equation}
Here, $M$ relates the field amplitudes at the left and
right leads of the bus waveguide.
Its explicit elements and detailed derivation are
provided in Appendix~A.

\subsection{Phase diagram}
Inspection of $M$ reveals that
$M_{22}=M_{11}^{*}$, $M_{12}^{*}=-M_{12}$, and
$M_{21}^{*}=-M_{21}$.
In addition, $\det M=1$ ensures a reciprocal transmission.
These relations bring
$M^{-1}=M^{*}$, demonstrating the effective
$\mathcal{PT}$ symmetry of our studying system.
As a result, these constraints allow us to parameterize $M$
following Ref.~\cite{Ge2012,Lee2021}.

We consider the scattering matrix $S$ through
\begin{equation}
\begin{bmatrix}
C^R\\
B^L
\end{bmatrix}
=S
\begin{bmatrix}
A^L\\
D^R
\end{bmatrix},
\qquad
S=\frac{1}{M_{22}}
\begin{bmatrix}
1 & M_{12}\\
-M_{21} & 1
\end{bmatrix}.
\end{equation}
We then write
$M_{12}=i\mu$ and $M_{21}=i\nu$, where $\mu$ and $\nu$
are real, and define $\eta=\mu\nu$.
The corresponding scattering eigenvalues and  right scattering eigenvectors are $s_{\pm}=\frac{1\pm\sqrt{\eta}}{M_{22}}$ and $\psi_{\pm}= \left[i\mu, \pm\sqrt{\eta}\right]^T$, respectively.
The components of $\boldsymbol{\psi}_{\pm}$ specify
the relative amplitudes and phases of the incident
fields $\left[A^L,D^R\right]^T$.
By analyzing scattering eigenvalues and eigenvectors, we can state that the regimes $\eta<0$ and $0<\eta<1$ correspond to the
$\mathcal{PT}$-symmetric and $\mathcal{PT}$-symmetry broken
phases, respectively, while the corresponding reciprocal transmission coefficient is defined as $\sigma\equiv \frac{C^R}{A^L}=\frac{B^L}{D^R}$.
The transmittance $T\equiv\vert \sigma\vert^2$ gives $T<1$ and $T>1$ in $\mathcal{PT}$-symmetric and $\mathcal{PT}$-symmetry broken
phases, respectively \cite{Ge2012,Lee2021}.
At $\eta=0$, provided that exactly one of $\mu$ and $\nu$
vanishes, the system reaches a
$\mathcal{PT}$-exceptional point with $T=1$, also known for symmetry breaking point \cite{Heiss2012,Chong2011}.
Our $\mathcal{PT}$-symmetry breaking point is a second-order exceptional point of the scattering matrix,
where both eigenvalues and eigenvectors coalesce \cite{Miri2019,Hodaei2017}.
The detailed derivation and classification are
provided in Appendix B.

Accordingly, Fig.~\ref{fig2} presents scattering phase
diagrams in the $(\tau,\alpha_R)$ plane for selected
values of the round-trip effective propagation phase
$\varphi\equiv 2\pi k_R\rho=[1,2,3,4,5,2\pi]$.
We write $k_R=n_{\mathrm{ring}}^{\mathrm{eff}}k_0$,
where $n_{\mathrm{ring}}^{\mathrm{eff}}$ is the effective
refractive index of two circulating modes for both rings,
$k_0=\frac{2\pi}{\lambda}$ is the vacuum wavenumber, and $\lambda$ is an incident wavelength.
In Fig.~\ref{fig2}, the yellow and red regions represent the
$\mathcal{PT}$-symmetric and $\mathcal{PT}$-broken
scattering phases, respectively, separated by
exceptional-point boundaries marked by green and
purple curves.
These diagrams provide a unified description of the
scattering behavior for systems satisfying the
assumptions of our model, irrespective of their
specific geometric and material parameters.

\section{Discussion}
\subsection{General unitary relation}
The reflection coefficients for an incidence from the right (loss-ring side) and left (gain-ring side) sides
are
$r_R\equiv
\left.\frac{C^R}{D^R}\right|_{A^L=0}
=\frac{M_{12}}{M_{22}}$ and 
$r_L\equiv
\left.\frac{B^L}{A^L}\right|_{D^R=0}
=-\frac{M_{21}}{M_{22}}$, respectively.
The corresponding reflectances are
$R_R=|r_R|^2$ and $R_L=|r_L|^2$.
Combining the parametrization with reciprocity and
$\mathcal{PT}$ symmetry yields the generalized
unitarity relation, $r_Rr_L=\sigma^2(1-\frac{1}{\vert\sigma\vert^2})$, already proposed in ~\cite{Ge2012}.
Taking the squared modulus yields the generalized
conservation relation $R_LR_R=(T-1)^2$ for reciprocal,
one-dimensional $\mathcal{PT}$-symmetric two-port systems.

\subsection{Uni-directional reflectionless, A self-dual spectral singularity, and transmission extinction}
To have a reflectionless at the right side, it should meet  $r_R=0$, yielding
\begin{equation}
\tau=\sqrt{2\alpha_R\cos[\varphi]-\alpha_R^2}.
\end{equation}
The generalized conservation relation further implies
$T=1$.
Since $\tau$ is real and bound by $0<\tau<1$, the allowed
attenuation factor must additionally satisfy $\alpha_R<2\cos\varphi$.
In Fig.~\ref{fig2}, the corresponding parameter combinations of $(\tau,\alpha_R,\varphi)$
satisfying $r_R=0$ are indicated by the green curves, which coincide with the
exceptional-point boundaries, as expected.

Similarly, a reflectionlessness for incidence from the
left requires $r_L=0$,  yielding  
\begin{equation}
\tau=\frac{1}{\alpha_R}\sqrt{2\alpha_R\cos[\varphi]-1}.
\end{equation}
The generalized conservation relation again implies
$T=1$.
For the requirement of real $\tau$ and $0<\tau<1$, it demands
$\cos\varphi>\frac{1}{2\alpha_R}$.
This condition is achievable only for $\alpha_R>1/2$.
The purple curves in Fig.~\ref{fig2} indicate $r_L=0$ which coincide with the exceptional-point boundaries
and occur only for $\alpha_R>1/2$, as expected.

      \begin{figure}[ht]
 \centering
\includegraphics[width=0.5\textwidth]{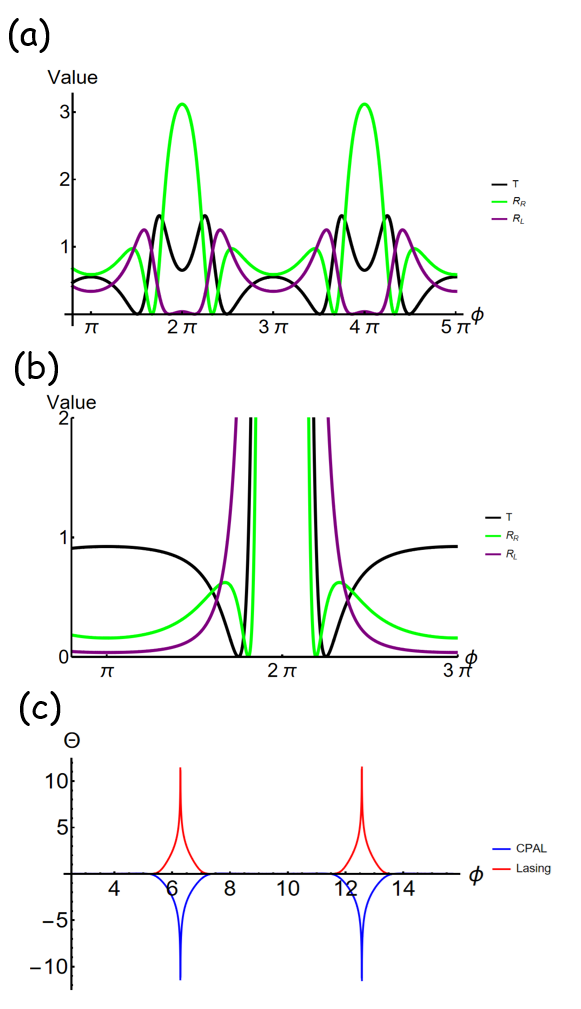}
\caption{Transmittance $T$ (black), right-incidence reflectance
$R_R$ (green), and left-incidence reflectance $R_L$
(purple) as functions of the round-trip propagation
phase $\varphi$. (a) For $\tau=0.5$ and $\alpha_R=0.6$, transmission
extinction and unidirectional reflectionlessness from
either side occur at distinct phases.
(b) For $\alpha_R=0.35$ and $\tau\approx0.6731$,
CPAL occurs at the resonances $\varphi=2m\pi$.
Each CPAL point is symmetrically flanked by right-incidence
reflectionlessness and transmission extinction on
both sides.
(c) Coherent perfect absorption (blue) and lasing
(red) under the corresponding excitation conditions
for the same parameters as in (b).
These scattering features recur with a period of
$2\pi$ in $\varphi$.}
\label{fig3}
\end{figure}

   \begin{figure*}[ht]
 \centering
\includegraphics[width=0.9\textwidth]{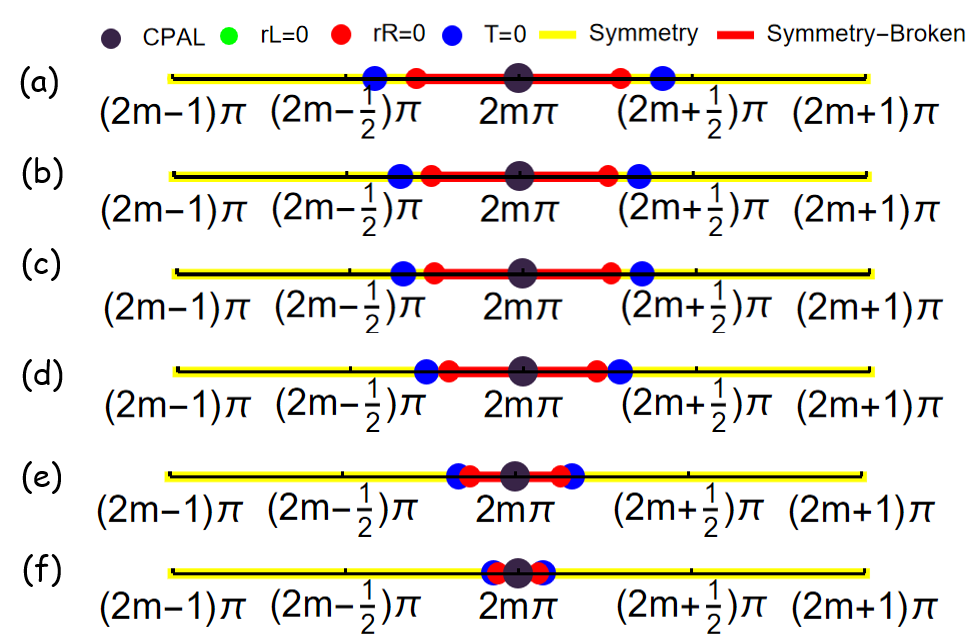}
\caption{Scattering features as functions of the round-trip
propagation phase $\varphi$ under the CPAL parameter
constraint, with (a) $\alpha_R=0.1$,
(b) $\alpha_R=0.2$, (c) $\alpha_R=0.3$,
(d) $\alpha_R=0.5$, (e) $\alpha_R=0.7$, and
(f) $\alpha_R=0.8$.
Dark black, red, and blue dots indicate CPAL at
$\varphi=2m\pi$, right-incidence reflectionlessness
($r_R=0$), and transmission extinction ($T=0$),
respectively.
Yellow and red bars denote the
$\mathcal{PT}$-symmetric and
$\mathcal{PT}$-broken phases, respectively.
Left-incidence reflectionlessness ($r_L=0$)
is absent under the CPAL parameter constraint.}
\label{fig4}
\end{figure*}

We further examine whether bidirectional reflectionlessness,
$r_R=0$ and $r_L=0$, can occur.
Combining the two reflectionless conditions yields $
(1-\tau^2)(1-\alpha_R^2)=0.$
For $0<\tau<1$, this requires $\alpha_R=1$, with
$\tau=\sqrt{2\cos\varphi-1}.$
Thus, bidirectional reflectionlessness is excluded
in the balanced gain--loss regime considered here, since $0<\alpha_R<1$.

Coherent perfect absorption (CPA) requires $M_{11}=0$,
which gives $
\alpha_R+\tau^2e^{-i\varphi}
\left[-1+\alpha_R(e^{-i\varphi}-\alpha_R)\right]=0.$
Similarly, the lasing threshold requires $M_{22}=0$,
yielding $
\alpha_R+\tau^2e^{i\varphi}
\left[-1+\alpha_R(e^{i\varphi}-\alpha_R)\right]=0$.
$\mathcal{PT}$ symmetry intrinsically imposes
$M_{11}=M_{22}^{*}$, as a result of their zero solutions being simultaneously satisfied to give CPAL.
Since $0<\alpha_R<1$ and $0<\tau<1$, the resulting CPAL conditions are
\begin{equation}
\label{eq:CPAL-condition}
\begin{cases}
\varphi=2m\pi,
& m=1,2,3,\ldots,\\[4pt]
\displaystyle
\tau^2=\frac{\alpha_R}{1+\alpha_R^2-\alpha_R}.
\end{cases}
\end{equation}
Thus, CPAL occurs at a whispering-gallery resonance
when the through-coupling coefficient and the
round-trip attenuation factor satisfy
Eq.~\eqref{eq:CPAL-condition}.
Consequently, at the CPAL point, both reflectances
$R_R$ and $R_L$ and the transmittance $T$ diverge under  a single-sided
excitation.
The corresponding parameter combinations are indicated
by the blue curve in Fig.~\ref{fig2}(f).

We next consider transmission extinction, i.e., $\sigma=0$,
which occurs when
\begin{equation}
\begin{aligned}
&\tau^4
+(-1+6\tau^2-3\tau^4+2\tau^6)\alpha_R^2
+\tau^4\alpha_R^4\\
&\quad
+2\tau^2\alpha_R
\bigl[
-(1+\tau^2)(1+\alpha_R^2)\cos\varphi
+\alpha_R\cos(2\varphi)
\bigr]=0,
\end{aligned}
\end{equation}
provided that the numerator of $\sigma$ remains nonzero.
A transmission extinction ($T=0$) lies in the
$\mathcal{PT}$-symmetric phase,  supporting by the corresponding parameter combinations with dashes black lines in  Fig.~\ref{fig2}(f).
With $T=0$, the generalized conservation relation yields $R_LR_R=1$,
so that the left and right reflectances form a reciprocal pair.

Although unidirectional reflectionlessness, transmission
extinction, and CPAL cannot occur at the same incident
wavelength, suitably chosen system parameters allow these phenomena to occur at distinct values of the
round-trip propagation phase $\varphi$, corresponding
to different incident wavelengths $\lambda$.
As an example, Fig.~\ref{fig3}(a) shows the scattering
response for $\tau=0.5$ and $\alpha_R=0.6$.
Transmission extinction ($T=0$) occurs at
$\varphi\approx4.74$, $7.83$, $11.02$, and $14.11$.
Right-incidence reflectionlessness ($R_R=0$) occurs at
$\varphi\approx5.25$, $7.32$, $11.50$, and $13.60$,
whereas left-incidence reflectionlessness ($R_L=0$)
occurs at $\varphi\approx5.85$, $6.70$, $12.13$,
and $12.99$.
Each set of scattering features repeats with a
phase period of $2\pi$.

We next consider a system satisfying the CPAL condition,
with $\alpha_R=0.35$ and $\tau\approx0.6731$,
as shown in Fig.~\ref{fig3}(b).
At the resonance $\varphi=2\pi$, the transmittance $T$
and both reflectances $R_R$ and $R_L$ diverge under
single-sided excitation in the linear model, indicating
the lasing threshold associated with CPAL.

To characterize the scattering response under coherent two-sided
excitation, we define the ratio of total outgoing to
total incoming power in the common bus waveguide as
$\Psi\equiv
\frac{|C^R|^2+|B^L|^2}
     {|A^L|^2+|D^R|^2},$
where the port amplitudes are consistently
power-normalized.
For incident light fields in the bus waveguide satisfying $D^R=M_{21}A^L$ with the appropriate relative amplitudes and
phases,
we then have
$\Psi_{\mathrm{CPA}}
=\frac{|M_{11}|^2}{1+|M_{21}|^2},$
which vanishes at the CPAL point, where $M_{11}=0$.
The numerical calculation is verified in a blue line of Fig.~\ref{fig3}(c).
For comparison, equal-amplitude, in-phase excitation,
$D^R=A^L$, gives 
$\Psi_{\mathrm{equal}}
=\frac{|1+M_{12}|^2+|1-M_{21}|^2}
       {2|M_{22}|^2}$, as shown in a red line of Fig.~\ref{fig3}(b).
Unless the excitation satisfies the CPA input condition,
any deviation leads to $\Psi$ amplified at the CPAL point.

Interestingly, Fig.~\ref{fig3}(b) shows transmission
extinction ($T=0$) and right-incidence reflectionlessness
($R_R=0$) near the CPAL point.
To understand these accompanying scattering features,
we examine the transmission and reflection coefficients
under the CPAL parameter constraint specified in
Eq.~\eqref{eq:CPAL-condition}.

Substituting this constraint into the condition $r_R=0$
yields$
\varphi=2m\pi\pm\varphi_R$,
$\qquad m\in\mathbb{Z},$
where
\begin{equation}
\varphi_R=
\arccos\left[
\frac{\alpha_R^3-\alpha_R^2+\alpha_R+1}
{2(\alpha_R^2-\alpha_R+1)}
\right].
\end{equation}
For $0<\alpha_R<1$, we give $
0<\varphi_R<\frac{\pi}{3}.$
Thus, right-incidence reflectionless points occur
symmetrically on either side of each CPAL resonance,
with phase offsets smaller than $\pi/3$.

Under the CPAL parameter constraint, the condition
$r_L=0$ requires
$
\cos\varphi_L=
\frac{\alpha_R^3+\alpha_R^2-\alpha_R+1}
{2(\alpha_R^3-\alpha_R^2+\alpha_R)}$.
However, through analysis, we can discover that for $0<\alpha_R<1$, no real propagation phase satisfies $r_L=0$.
This remarkable result reveals that the CPAL parameter constraint
excludes left-incidence reflectionlessness (the gain ring side), in contrast
to the example with $R_L=0$ in Fig.~\ref{fig3}(a), whose the parameters
do not satisfy this constraint.

We next examine transmission extinction ($T=0$)
under the CPAL parameter constraint.
The corresponding transmission extinction condition
reduces to
$
\varphi_T=\arccos\left[\frac{B\pm\sqrt{B^2-4AC}}{2A}\right]$
where 
$A=4t^2\alpha_R^2$, 
$B=2t^2\alpha_R(1+t^2)(1+\alpha_R^2)$, and $C=t^4+(-1+6t^2-3t^4+2t^6)\alpha_R^2-2t^2\alpha_R^2+t^4\alpha_R^4$.
However, for $0<\alpha_R<1$, only the root
\begin{equation}
\varphi_T=\arccos\left[\frac{B-\sqrt{B^2-4AC}}{2A}\right]
\end{equation}
yields a real propagation phase.
The transmission extinction points are consequently
located at
\begin{equation}
\varphi=2m\pi\pm\varphi_T,\qquad m\in\mathbb{Z}.
\end{equation}

The round-trip phase separation between the two transmission extinction points
bracketing a CPAL resonance is thus
\begin{equation}\label{Eq:variphi}
\Delta\varphi_T
=2\varphi_T
=2\arccos\left[
\frac{B-\sqrt{B^2-4AC}}{2A}
\right].
\end{equation}

Finally, Fig.~\ref{fig3} summarizes the  locations of transmission extinction,
right-incidence reflectionlessness, and CPAL,
together with the associated $\mathcal{PT}$ scattering phases,
as functions of $\varphi$ under the CPAL parameter constraint.
The round-trip attenuation factors are
(a) $\alpha_R=0.1$, (b) $\alpha_R=0.2$, (c) $\alpha_R=0.3$,
(d) $\alpha_R=0.5$, (e) $\alpha_R=0.7$, and (f) $\alpha_R=0.8$.

As $\varphi$ increases through each resonance
$\varphi=2m\pi$, these systems exhibit a common
sequence of scattering features:
transmission extinction, right-incidence
reflectionlessness, CPAL, right-incidence
reflectionlessness, and transmission extinction.
This sequence repeats with a phase period of $2\pi$.
Furthermore, the phase separation $\Delta\varphi_T$
decreases as $\alpha_R$ approaches unity, corresponding
to weaker round-trip amplification and attenuation,
consistent with Eq.~\eqref{Eq:variphi}.

  \begin{figure*}[ht]
 \centering
\includegraphics[width=1\textwidth]{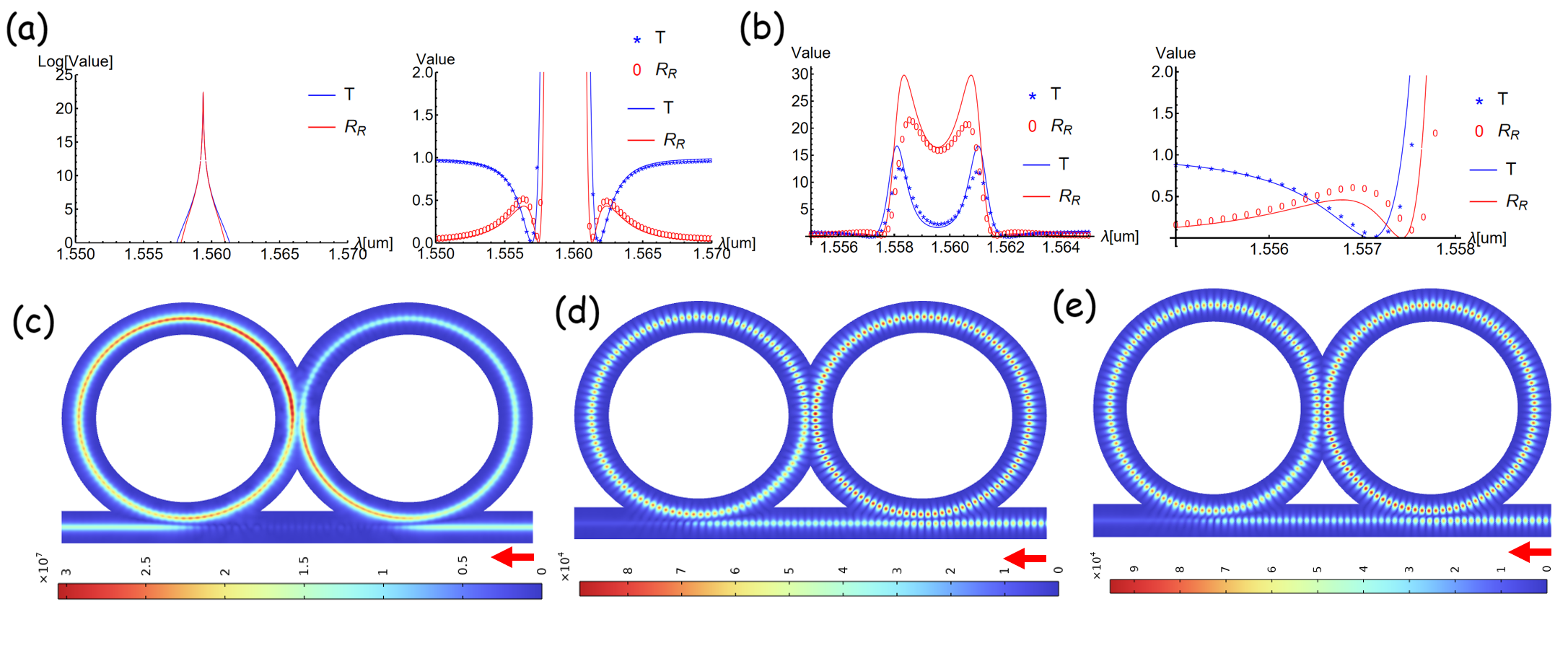}
\caption{Analytical and stimulating scattering responses of two mutually coupled rings externally coupled to a bus waveguide.
(a) Transmittance $T$ (blue) and right-incidence
reflectance $R_R$ (red) for
$\tau\approx0.858$ and $\alpha_R\approx0.55$, satisfying the CPAL parameter constraint.
The left panel shows the resonant divergence near
$\lambda\approx1.558\,\mu\mathrm{m}$ predicted by our analytical
model, while the right panel highlights the accompanying
transmission extinction and right-incidence
reflectionlessness. Here blue crosses and red circles are obtained by the COMSOL Multiphysics simulation.
(b) Corresponding spectra for $\tau=0.89$ and
$\alpha_R=0.6875$, which do not satisfy the CPAL
parameter constraint.
The right panel shows an enlarged view of transmission
extinction and right-incidence reflectionlessness.
Solid curves represent analytical results, while
blue stars and red circles denote COMSOL Multiphysics
results for $T$ and $R_R$, respectively.
(c)--(e) Simulated electric-field magnitude distributions
under right incidence:
(c) near the CPAL resonance in (a),
(d) at transmission extinction in (a), and
(e) at transmission extinction in (b).
Red arrows indicate the incident-wave direction.}
\label{fig5}
\end{figure*}

\section{Electromagnetic simulations}
To verify our analytical results, we perform
electromagnetic simulations using COMSOL Multiphysics.
We first consider a single ring resonator coupled
to a straight bus waveguide, with a lossless core
material in the ring.
This configuration is used to identify the
whispering gallery resonances and extract the
through-coupling coefficient $\tau$ and round-trip
attenuation factor $\alpha_R$.
The resonance condition is determined by the
round-trip propagation phase, whereas critical
coupling additionally requires the round-trip
attenuation factor to match the magnitude of
the through-coupling coefficient.

We distinguish the material refractive index from
the effective refractive index of a guided mode.
The latter determines the propagation phase in
the ring and bus waveguides.
Both waveguides have the same width and core
and cladding materials.
Our numerical results indicate that the real parts
of their effective refractive indices are approximately
equal for the configurations considered here.
Even with lossless constituent materials, the ring
can exhibit attenuation through bending-induced
radiation and other leakage loss.
Such attenuation can be represented by an imaginary
part of the effective propagation constant and incorporated
into the round-trip field-amplitude factor $\alpha_R$.

The coupling coefficient $\tau$ is controlled primarily
by the gap between adjacent waveguides.
Within the coupling regime considered here, reducing
the gap enhances cross coupling and decreases
the through-coupling coefficient $\tau$.
The round-trip factors $\alpha_R$ and $\alpha_L$
are controlled primarily by introducing material
loss and gain, respectively.
Increasing the loss decreases $\alpha_R$, whereas
increasing the gain increases $\alpha_L$.
Although these adjustments can also affect the modal
indices and coupling, our numerical results indicate
that the coupling and round-trip amplitude factors
can be tuned approximately independently over the
parameter range considered.

The parameter extraction and inverse-design procedure
for the single-ring configuration are detailed in
Appendix~\ref{app:design}.
Using the calibrated components, we then construct
the two mutually coupled rings, each coupled to
the common bus waveguide, and verify the resulting
scattering response.

To test the predicted CPAL response, we consider
$\tau\approx0.858$ and $\alpha_R\approx0.55$,
which lie close to the CPAL parameter condition.
The left panel of Fig.~\ref{fig5}(a) shows the
analytical spectra, with a pronounced divergence
of the transmittance $T$ and right-incidence
reflectance $R_R$ at the resonance near
$\lambda_0=1.559\,\mu\mathrm{m}$.
The right panel compares the analytical results
(solid curves) with electromagnetic simulations
performed using COMSOL Multiphysics
(blue stars for $T$ and red circles for $R_R$).
Material dispersion is neglected in these simulations.
The numerical results reproduce the main spectral
features, including transmission extinction and
right-incidence reflectionlessness on either side
of the central resonance.

Figure~\ref{fig5}(c) shows the simulated electric-field
magnitude near the CPAL resonance under excitation
from the right.
The field is strongly enhanced within the rings,
reaching values on the order of
$10^7\,\mathrm{V/m}$ of the electric field.

Figure~\ref{fig5}(d) shows the field distribution
at $\lambda_0=1.5569\,\mu\mathrm{m}$, corresponding
to transmission extinction.
The strongly suppressed field in the left output
segment of the bus waveguide supports the predicted
vanishing transmittance.
This extinction results from destructive interference
between the directly transmitted field and the fields
coupled back into the bus waveguide from the rings.
Compared with Fig.~\ref{fig5}(c), the pronounced
interference fringes on the right side indicate
a substantial reflected field.
At this wavelength, the simulated field magnitude
in the loss ring can exceed that in the gain ring.

For comparison, we consider a second system with
$\tau=0.89$ and $\alpha_R=0.6875$, which does not
satisfy the CPAL parameter condition.
Figure~\ref{fig5}(b) compares the analytical spectra
with the COMSOL results using the same line and
symbol conventions as in Fig.~\ref{fig5}(a).
Unlike the CPAL case, both $T$ and $R_R$ exhibit
a local minimum near the central resonance at
$\lambda_0\approx1.559\,\mu\mathrm{m}$,
flanked by two peaks.
Transmission extinction and right-incidence
reflectionlessness also occur away from the central
resonance, as highlighted in the right panel.
The simulated extinction and reflectionless
wavelengths are approximately
$1.5571\,\mu\mathrm{m}$ and
$1.55743\,\mu\mathrm{m}$, respectively.
The field distribution at the extinction transmission wavelength,
shown in Fig.~\ref{fig5}(e), again exhibits strongly
suppressed transmission into the left output segment,
consistent with the analytical prediction.

The simple ring--bus architecture considered here
builds on established silicon photonic components
compatible with complementary metal--oxide--semiconductor
(CMOS) fabrication technologies~\cite{Bogaerts2012}.
Active microring devices have also been demonstrated
through hybrid III--V/silicon integration~\cite{Liang2009}.
These advances suggest a route toward realizing
the predicted scattering phenomena in silicon
photonic integrated circuits.

\section{Conclusion}
We have analytically established an effective
$\mathcal{PT}$ symmetry in two mutually coupled ring
resonators.
Combining this symmetry with reciprocity, we developed a parametrized phase diagram that classifies the
scattering phases and identifies transmission extinction
and reflectionlessness for incidence from either side.
CPAL occurs at a whispering-gallery resonance when
the through-coupling coefficient and the round-trip attenuation factor satisfy a specific relation.

Under this CPAL parameter constraint, all systems exhibit a common sequence of scattering features
around each resonance: transmission extinction,
loss-side reflectionlessness, CPAL, loss-side
reflectionlessness, and transmission extinction.
This sequence repeats periodically in the round-trip propagation phase.
We derived the phase separation between the extinction points bracketing each CPAL resonance, which narrows as the round-trip gain and loss factors become weaker.
We also showed that gain-side reflectionlessness is
excluded under the CPAL parameter constraint but can occur for other parameter combinations.
Electromagnetic simulations using COMSOL Multiphysics reproduce the principal analytical scattering features.
Together with established ring-resonator integration technologies, our work provides design guidelines for exploring CPAL and its accompanying scattering features in silicon photonic platforms.

\section{Appendix A:Derivation of the transfer matrix}
\label{app:transfer}

At the inter-ring coupling region, the counterclockwise mode
in the left ring couples to the clockwise mode in the right
ring according to
\begin{equation}
\begin{pmatrix}
v^{L}(2^{+})\\
u^{R}(2^{+})
\end{pmatrix}
=
\mathcal{T}
\begin{pmatrix}
v^{L}(2^{-})\\
u^{R}(2^{-})
\end{pmatrix}.
\end{equation}
Using $\tau^2+\kappa^2=1$ and rearranging these relations,
we obtain
\begin{equation}
\begin{pmatrix}
u^{R}(2^{+})\\
u^{R}(2^{-})
\end{pmatrix}
=
M_1
\begin{pmatrix}
v^{L}(2^{+})\\
v^{L}(2^{-})
\end{pmatrix},
\end{equation}
where
\begin{equation}
M_1=
\frac{1}{i\kappa}
\begin{pmatrix}
\tau & -1\\
1 & -\tau
\end{pmatrix}.
\label{eq:M1}
\end{equation}

Next, we consider the coupling between the clockwise mode
in the right ring and the left-propagating mode in the bus
waveguide:
\begin{equation}
\begin{pmatrix}
u^{R}(3^{+})\\
B^{R}
\end{pmatrix}
=
\mathcal{T}
\begin{pmatrix}
u^{R}(3^{-})\\
D^{R}
\end{pmatrix}.
\end{equation}
Rearranging this equation gives
\begin{equation}
\begin{pmatrix}
D^{R}\\
B^{R}
\end{pmatrix}
=
\frac{1}{i\kappa}
\begin{pmatrix}
1 & -\tau\\
\tau & -1
\end{pmatrix}
\begin{pmatrix}
u^{R}(3^{+})\\
u^{R}(3^{-})
\end{pmatrix}.
\end{equation}

Propagation along the right ring relates the field amplitudes
at coupling regions $2$ and $3$ through
\begin{align}
u^{R}(3^{-})
&=
\alpha_R^{3/4}
e^{i3\pi k_R\rho/2}
u^{R}(2^{+}),\\
u^{R}(2^{-})
&=
\alpha_R^{1/4}
e^{i\pi k_R\rho/2}
u^{R}(3^{+}),
\end{align}
where $\rho$ is the ring radius and $\alpha_R$ is the
round-trip field-amplitude attenuation factor. The fractional
powers of $\alpha_R$ assume uniformly distributed attenuation
along the ring.

Substituting these propagation relations yields
\begin{equation}
\begin{pmatrix}
D^{R}\\
B^{R}
\end{pmatrix}
=
M_2
\begin{pmatrix}
u^{R}(2^{-})\\
u^{R}(2^{+})
\end{pmatrix},
\end{equation}
with
\begin{equation}
M_2=
\frac{1}{i\kappa}
\begin{pmatrix}
\alpha_R^{-1/4}e^{-i\pi k_R\rho/2}
&
-\tau\alpha_R^{3/4}e^{i3\pi k_R\rho/2}
\\
\tau\alpha_R^{-1/4}e^{-i\pi k_R\rho/2}
&
-\alpha_R^{3/4}e^{i3\pi k_R\rho/2}
\end{pmatrix}.
\label{eq:M2}
\end{equation}

Propagation in the left ring gives
\begin{align}
v^{L}(2^{-})
&=\alpha_L^{1/4}e^{i\pi k_L\rho/2}v^{L}(1^{+}),\\
v^{L}(1^{-})
&=\alpha_L^{3/4}e^{i3\pi k_L\rho/2}v^{L}(2^{+})
\end{align}
where $\alpha_L$ is defined as the round-trip field-amplitude amplification factor and assumed to be uniformly distributed along the ring. 　

These relations can be written as
\begin{equation}
\begin{pmatrix}
v^{L}(2^{-})\\
v^{L}(2^{+})
\end{pmatrix}
=M_3
\begin{pmatrix}
v^{L}(1^{-})\\
v^{L}(1^{+})
\end{pmatrix},
\end{equation}
where
\begin{equation}
M_3=
\begin{pmatrix}
0 & \alpha_L^{1/4}e^{i\pi k_L\rho/2}\\
\alpha_L^{-3/4}e^{-i3\pi k_L\rho/2} & 0
\end{pmatrix}.
\end{equation}

At the coupling region between the left ring and the bus
waveguide, the counterclockwise mode satisfies
\begin{equation}
\begin{pmatrix}
v^{L}(1^{+})\\
C^L
\end{pmatrix}
=\mathcal{T}
\begin{pmatrix}
v^{L}(1^{-})\\
A^L
\end{pmatrix}.
\end{equation}
Rearranging this relation yields
\begin{equation}
\begin{pmatrix}
A^L\\
C^L
\end{pmatrix}
=M_4
\begin{pmatrix}
v^{L}(1^{-})\\
v^{L}(1^{+})
\end{pmatrix},
\qquad
M_4=\frac{1}{i\kappa}
\begin{pmatrix}
-\tau & 1\\
-1 & \tau
\end{pmatrix}.
\end{equation}

To account for the ordering of the field amplitudes, we
introduce the permutation matrix
\begin{equation}
J=
\begin{pmatrix}
0 & 1\\
1 & 0
\end{pmatrix}.
\end{equation}
Combining the coupling and propagation relations gives
\begin{equation}\label{eq:P}
\begin{pmatrix}
D^R\\
B^R
\end{pmatrix}
=P
\begin{pmatrix}
A^L\\
C^L
\end{pmatrix},
\qquad
P=M_2JM_1JM_3M_4^{-1}.
\end{equation}
Since $JM_1J=M_4$, this expression simplifies to
\begin{equation}
P=M_2M_4M_3M_4^{-1}.
\end{equation}

\subsubsection{Balanced amplification-dissipation and equality of propagation constants}

We impose reciprocal round-trip amplitude factors and equal
propagation constants,
\begin{equation}
\alpha_L\alpha_R=1,
\qquad
k_L=k_R.
\end{equation}
Defining $\alpha=\alpha_R$ and
$\varphi=2\pi k_R\rho$, we obtain
\begin{equation}
P=
\begin{pmatrix}
P_{11} & P_{12}\\
P_{21} & P_{22}
\end{pmatrix},
\end{equation}
with
\begin{align}
P_{11}
&=\frac{i\tau}{\kappa^3\sqrt{\alpha}}
  \left(1+\tau^2\alpha^2-2\alpha\cos\varphi\right),\\
P_{12}
&=\frac{i}{\kappa^3\sqrt{\alpha}}
  \left[
  \alpha e^{-i\varphi}
  +\tau^2\alpha e^{i\varphi}
  -\tau^2(1+\alpha^2)
  \right],\\
P_{21}
&=\frac{i}{\kappa^3\sqrt{\alpha}}
  \left[
  \tau^2(1+\alpha^2)
  -\alpha e^{i\varphi}
  -\tau^2\alpha e^{-i\varphi}
  \right],\\
P_{22}
&=-\frac{i\tau}{\kappa^3\sqrt{\alpha}}
  \left(\tau^2+\alpha^2-2\alpha\cos\varphi\right),
\end{align}
where $\tau^2+\kappa^2=1$ has been used.

\subsubsection{Coupling between the clockwise mode in the left  ring and the counterclockwise mode in the right ring}
At the inter-ring coupling region, the clockwise mode in
the left ring couples to the counterclockwise mode in the
right ring according to
\begin{equation}
\begin{pmatrix}
v^{R}(2^{+})\\
u^{L}(2^{+})
\end{pmatrix}
=\mathcal{T}
\begin{pmatrix}
v^{R}(2^{-})\\
u^{L}(2^{-})
\end{pmatrix}.
\end{equation}
Rearranging this relation gives
\begin{equation}
\begin{pmatrix}
v^{R}(2^{+})\\
v^{R}(2^{-})
\end{pmatrix}
=\frac{1}{i\kappa}
\begin{pmatrix}
\tau & -1\\
1 & -\tau
\end{pmatrix}
\begin{pmatrix}
u^{L}(2^{+})\\
u^{L}(2^{-})
\end{pmatrix}.
\end{equation}

Propagation in the left ring gives
\begin{align}
u^{L}(1^{-})
&=\alpha_L^{1/4}e^{i\pi k_L\rho/2}
  u^{L}(2^{+}),\\
u^{L}(2^{-})
&=\alpha_L^{3/4}e^{i3\pi k_L\rho/2}
  u^{L}(1^{+}).
\end{align}
Combining these propagation and coupling relations yields
\begin{equation}
\begin{pmatrix}
v^{R}(2^{+})\\
v^{R}(2^{-})
\end{pmatrix}
=N_1
\begin{pmatrix}
u^{L}(1^{+})\\
u^{L}(1^{-})
\end{pmatrix},
\end{equation}
where
\begin{equation}
N_1=\frac{1}{i\kappa}
\begin{pmatrix}
-\alpha_L^{3/4}e^{i3\pi k_L\rho/2}
&
\tau\alpha_L^{-1/4}e^{-i\pi k_L\rho/2}
\\
-\tau\alpha_L^{3/4}e^{i3\pi k_L\rho/2}
&
\alpha_L^{-1/4}e^{-i\pi k_L\rho/2}
\end{pmatrix}.
\end{equation}

At the coupling region between the left ring and the bus
waveguide, the clockwise mode satisfies
\begin{equation}
\begin{pmatrix}
u^{L}(1^{+})\\
B^L
\end{pmatrix}
=\mathcal{T}
\begin{pmatrix}
u^{L}(1^{-})\\
D^L
\end{pmatrix}.
\end{equation}
This relation can be rewritten as
\begin{equation}
\begin{pmatrix}
B^L\\
D^L
\end{pmatrix}
=N_2
\begin{pmatrix}
u^{L}(1^{+})\\
u^{L}(1^{-})
\end{pmatrix},
\qquad
N_2=\frac{1}{i\kappa}
\begin{pmatrix}
\tau & -1\\
1 & -\tau
\end{pmatrix}.
\end{equation}
Consequently,
\begin{equation}
\label{eq:left-clockwise}
\begin{pmatrix}
B^L\\
D^L
\end{pmatrix}
=N_2N_1^{-1}
\begin{pmatrix}
v^{R}(2^{+})\\
v^{R}(2^{-})
\end{pmatrix}.
\end{equation}

Similarly, at the coupling region between the right ring
and the bus waveguide, the counterclockwise mode satisfies
\begin{equation}
\begin{pmatrix}
v^{R}(3^{+})\\
C^R
\end{pmatrix}
=\mathcal{T}
\begin{pmatrix}
v^{R}(3^{-})\\
A^R
\end{pmatrix}.
\end{equation}
Rearranging this relation gives
\begin{equation}
\begin{pmatrix}
C^R\\
A^R
\end{pmatrix}
=N_2
\begin{pmatrix}
v^{R}(3^{+})\\
v^{R}(3^{-})
\end{pmatrix}.
\end{equation}

Propagation in the right ring gives
\begin{align}
v^{R}(2^{-})
&=\alpha_R^{3/4}e^{i3\pi k_R\rho/2}
  v^{R}(3^{+}),\\
v^{R}(3^{-})
&=\alpha_R^{1/4}e^{i\pi k_R\rho/2}
  v^{R}(2^{+}).
\end{align}
We therefore obtain
\begin{equation}
\begin{pmatrix}
C^R\\
A^R
\end{pmatrix}
=N_3
\begin{pmatrix}
v^{R}(2^{+})\\
v^{R}(2^{-})
\end{pmatrix},
\end{equation}
where
\begin{equation}
N_3=\frac{1}{i\kappa}
\begin{pmatrix}
-\alpha_R^{1/4}e^{i\pi k_R\rho/2}
&
\tau\alpha_R^{-3/4}e^{-i3\pi k_R\rho/2}
\\
-\tau\alpha_R^{1/4}e^{i\pi k_R\rho/2}
&
\alpha_R^{-3/4}e^{-i3\pi k_R\rho/2}
\end{pmatrix}.
\end{equation}

Combining this result with Eq.~\eqref{eq:left-clockwise}
yields
\begin{equation}
\label{eq:Q-relation}
\begin{pmatrix}
B^L\\
D^L
\end{pmatrix}
=Q
\begin{pmatrix}
C^R\\
A^R
\end{pmatrix},
\qquad
Q=N_2N_1^{-1}N_3^{-1}.
\end{equation}

Again, under the conditions
\begin{equation}
\alpha_L\alpha_R=1,
\qquad
k_L=k_R.
\end{equation}

The matrix $Q$ in Eq.~\eqref{eq:Q-relation} then becomes
\begin{equation}
Q=
\begin{pmatrix}
Q_{11} & Q_{12}\\
Q_{21} & Q_{22}
\end{pmatrix},
\end{equation}
with
\begin{align}
Q_{11}
&=-\frac{i\tau}{\kappa^3\sqrt{\alpha}}
  \left(1+\tau^2\alpha^2-2\alpha\cos\varphi\right),\\
Q_{12}
&=\frac{i}{\kappa^3\sqrt{\alpha}}
  \left[
  \tau^2(1+\alpha^2)
  -\alpha e^{i\varphi}
  -\tau^2\alpha e^{-i\varphi}
  \right],\\
Q_{21}
&=\frac{i}{\kappa^3\sqrt{\alpha}}
  \left[
  \alpha e^{-i\varphi}
  +\tau^2\alpha e^{i\varphi}
  -\tau^2(1+\alpha^2)
  \right],\\
Q_{22}
&=\frac{i\tau}{\kappa^3\sqrt{\alpha}}
  \left(\tau^2+\alpha^2-2\alpha\cos\varphi\right),
\end{align}
where $\tau^2+\kappa^2=1$ has been used.

\subsection{Overall transfer matrix under balanced amplification and dissipation}
Rearranging Eq.(\ref{eq:P}), we obtain
\begin{align}
A^L
&=-\frac{P_{22}}{P_{21}}C^L
  +\frac{1}{P_{21}}B^R,\\
D^R
&=-\frac{\det P}{P_{21}}C^L
  +\frac{P_{11}}{P_{21}}B^R.
\end{align}
Similarly, rearranging Eq.(\ref{eq:Q-relation}) gives
\begin{align}
C^R
&=\frac{1}{Q_{21}}D^L
  -\frac{Q_{22}}{Q_{21}}A^R,\\
B^L
&=\frac{Q_{11}}{Q_{21}}D^L
  -\frac{\det Q}{Q_{21}}A^R.
\end{align}

Propagation through the bus waveguide between the two
ring--bus coupling regions introduces a real phase
$\theta$, such that
\begin{equation}
A^R=e^{i\theta}C^L,
\qquad
D^L=e^{i\theta}B^R.
\end{equation}
Substituting these relations yields
\begin{align}
\begin{pmatrix}
C^R\\
D^R
\end{pmatrix}
&=X
\begin{pmatrix}
C^L\\
B^R
\end{pmatrix},\\
\begin{pmatrix}
A^L\\
B^L
\end{pmatrix}
&=Y
\begin{pmatrix}
C^L\\
B^R
\end{pmatrix},
\end{align}
where
\begin{align}
X&=
\begin{pmatrix}
-\dfrac{Q_{22}}{Q_{21}}e^{i\theta}
&
\dfrac{e^{i\theta}}{Q_{21}}
\\[6pt]
-\dfrac{\det P}{P_{21}}
&
\dfrac{P_{11}}{P_{21}}
\end{pmatrix},\\
Y&=
\begin{pmatrix}
-\dfrac{P_{22}}{P_{21}}
&
\dfrac{1}{P_{21}}
\\[6pt]
-\dfrac{\det Q}{Q_{21}}e^{i\theta}
&
\dfrac{Q_{11}}{Q_{21}}e^{i\theta}
\end{pmatrix}.
\end{align}

Eliminating the intermediate field amplitudes $C^L$ and
$B^R$, we obtain the overall transfer relation
\begin{equation}
\begin{pmatrix}
C^R\\
D^R
\end{pmatrix}
=M
\begin{pmatrix}
A^L\\
B^L
\end{pmatrix},
\qquad
M=XY^{-1}.
\end{equation}

Under the conditions $\alpha_L\alpha_R=1$ and $k_L=k_R$,
we again use $\alpha=\alpha_R$ and
$\varphi=2\pi k_R\rho$.
For compactness, we define
\begin{align}
F&=1+\tau^2\alpha^2-2\alpha\cos\varphi,\\
G&=\tau^2+\alpha^2-2\alpha\cos\varphi,\\
H_{\pm}
&=\alpha e^{\pm i\varphi}
  +\tau^2\alpha e^{\mp i\varphi}
  -\tau^2(1+\alpha^2),\\
\mathcal{D}&=\tau^2FG-\kappa^6\alpha^2.
\end{align}
The elements of the overall transfer matrix are then
\begin{align}
\label{eq:overall-M11}
M_{11}
&=\frac{e^{i\theta}H_+^2}{\mathcal{D}},\\
M_{12}
&=\frac{2i\tau\kappa^3\sqrt{\alpha}\,G}
        {\mathcal{D}},\\
M_{21}
&=-\frac{2i\tau\kappa^3\alpha^{3/2}F}
         {\mathcal{D}},\\
\label{eq:overall-M22}
M_{22}
&=\frac{e^{-i\theta}H_-^2}{\mathcal{D}},
\end{align}
where $\tau^2+\kappa^2=1$ has been used.

All parameters $\tau$, $\kappa$, $\alpha$, $k_R$, $\rho$,
and $\theta$ are real, with $0<\alpha<1$.
Consequently, $H_-=H_+^*$, and the transfer matrix satisfies
\begin{equation}
M_{22}=M_{11}^*,
\qquad
M_{12}^*=-M_{12},
\qquad
M_{21}^*=-M_{21}.
\end{equation}
Together with reciprocity, i.e.,$\det M=1$, these relations imply
\begin{equation}
M^{-1}=M^*,
\end{equation}
demonstrating the  effective $\mathcal{PT}$ symmetry in our two mutually coupled ring system.

 \section{Appendix B: $\mathcal{PT}$ symmetry phase, symmetry breaking point, and $\mathcal{PT}$ symmetry broken phase} 
The scattering
matrix $S$ in our analysis is through
\begin{equation}
\begin{pmatrix}
C^R\\
B^L
\end{pmatrix}
=S
\begin{pmatrix}
A^L\\
D^R
\end{pmatrix},
\qquad
S=\frac{1}{M_{22}}
\begin{pmatrix}
1 & M_{12}\\
-M_{21} & 1
\end{pmatrix}.
\end{equation}
Since the off-diagonal elements of $M$ are purely
imaginary, we write
\begin{equation}
M_{12}=i\mu,
\qquad
M_{21}=i\nu,
\qquad
\eta=\mu\nu,
\end{equation}
where $\mu$ and $\nu$ are real.
Using $\det M=1$ and $M_{11}=M_{22}^{*}$, we obtain
\begin{equation}
|M_{22}|^2=1-\eta,
\qquad
T=\frac{1}{1-\eta}.
\end{equation}

The scattering eigenvalues are
\begin{equation}
\label{eq:scattering-eigenvalues}
s_{\pm}=\frac{1\pm\sqrt{\eta}}{M_{22}}.
\end{equation}
For $\mu\neq0$, the corresponding right eigenvectors
can be chosen as
\begin{equation}
\label{eq:scattering-eigenvectors}
\boldsymbol{\psi}_{\pm}
=
\begin{pmatrix}
i\mu\\
\pm\sqrt{\eta}
\end{pmatrix},
\qquad
S\boldsymbol{\psi}_{\pm}
=s_{\pm}\boldsymbol{\psi}_{\pm}.
\end{equation}
These eigenvectors specify the relative amplitudes and
phases of the two incident fields required to excite
each scattering eigen ports:
\begin{equation}
\frac{D^R}{A^L}
=\frac{\pm\sqrt{\eta}}{i\mu}.
\end{equation}

For $\eta<0$, both eigenvalues are unimodular,
$|s_+|=|s_-|=1$, corresponding to the
$\mathcal{PT}$-symmetric phase with $T<1$.
For $0<\eta<1$, their magnitudes satisfy
$|s_+||s_-|=1$, with $|s_+|>1$ and $|s_-|<1$.
This regime corresponds to the $\mathcal{PT}$- symmetric broken
phase with $T>1$, in which the two eigenvalues
exhibit amplification and attenuation, respectively.

At $\eta=0$, the eigenvalues coincide at
$s_+=s_-=1/M_{22}$, and $T=1$.
If exactly one of $\mu$ and $\nu$ vanishes, the
eigenvectors also coalesce, yielding a second-order
exceptional point.
For $\nu=0$ and $\mu\neq0$, the coalesced eigenvector is
$(1,0)^T$, corresponding to left incidence and
reflectionlessness from the left.
For $\mu=0$ and $\nu\neq0$, it is $(0,1)^T$,
corresponding to right incidence and reflectionlessness
from the right.
However, if $\mu=\nu=0$, $S=M_{22}^{-1}I$ supports
two independent eigenvectors and describes
bidirectional reflectionlessness rather than an
exceptional point.

 \section{Appendix C: Design principles for coupled ring resonators}\label{app:design}
 
   \begin{figure}[ht]
 \centering
\includegraphics[width=0.5\textwidth]{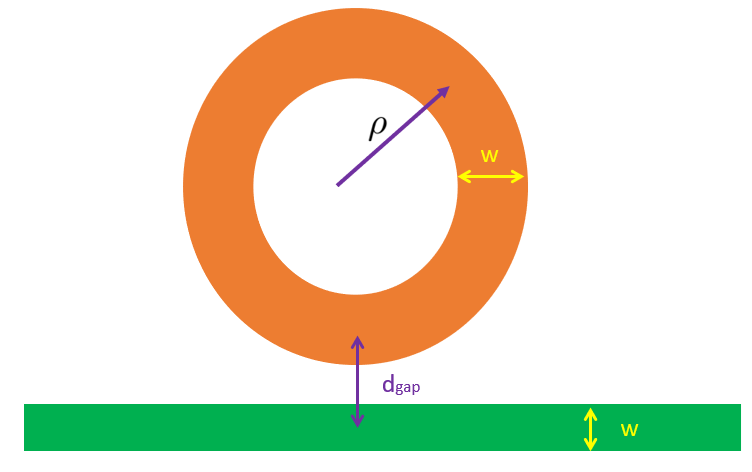}
\caption{Schematic of a single ring externally coupled to a bus waveguide.}
\label{app1}
\end{figure}

  \begin{figure*}[ht]
 \centering
\includegraphics[width=1\textwidth]{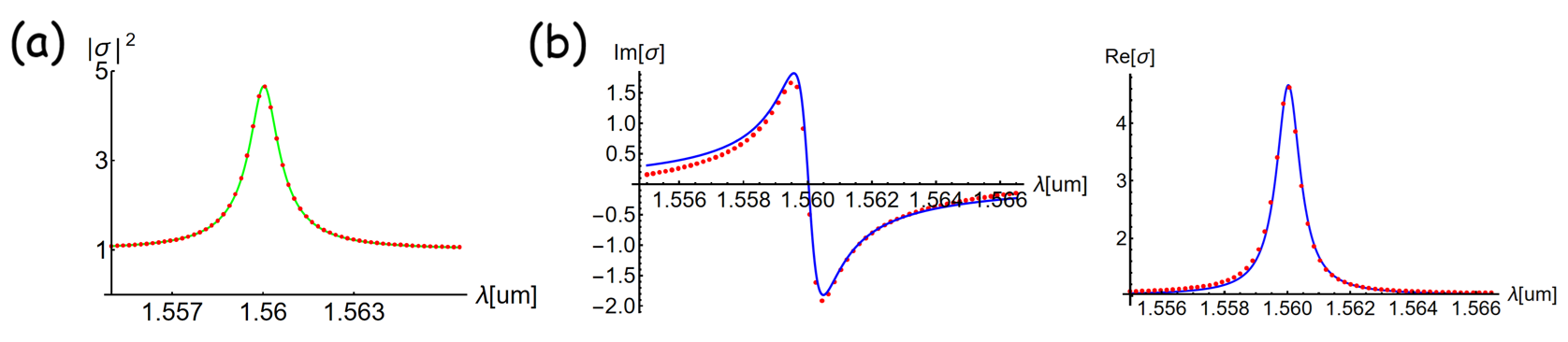}
\caption{Comparison of the analytical model (solid curves) and
COMSOL Multiphysics simulations (red dots) for the
complex transmission coefficient $\sigma$ of a single
ring resonator coupled to a bus waveguide:
(a) transmittance $|\sigma|^2$,
(b) imaginary part $\operatorname{Im}\left[\sigma\right]$, and
(c) real part $\operatorname{Re}\left[\sigma\right]$. }
\label{app2}
\end{figure*}

 We first consider a single ring resonator externally
coupled to a straight bus waveguide, as shown in
Fig.~\ref{app1}.
Both waveguides have a width of
$w=0.2\,\mu\mathrm{m}$, and the ring radius is
$\rho=6.2\,\mu\mathrm{m}$.
The edge-to-edge coupling gap, denoted by
$d_{\mathrm{gap}}$, is varied to obtain the desired
through-coupling coefficient $\tau$.
The core and cladding refractive indices are initially
set to $2.5$ and $1.5$, respectively, for both waveguides.

Using COMSOL Multiphysics, we calculate the complex
transmission coefficient $\sigma$ under single-sided
excitation over the wavelength range
$1.555$--$1.566\,\mu\mathrm{m}$.
The spectrum exhibits a whispering-gallery resonance
near $\lambda_0=1.56\,\mu\mathrm{m}$, accompanied by
a pronounced transmission minimum.
We then introduce material gain into the ring core
by setting
\begin{equation}
n_{\mathrm{ring}}=2.5-0.0051i,
\end{equation}
while retaining the remaining material and geometric
parameters.
The resulting transmission spectrum exhibits a
resonant peak near the same wavelength.

To extract the effective model parameters, we fit
the simulated complex transmission coefficient
using the single-ring expression
$\sigma=
\frac{\tau-\alpha e^{i\varphi}}
     {1-\tau\alpha e^{i\varphi}},
\qquad
\varphi=k_0n_{\mathrm{ring}}^{\mathrm{eff}}2\pi\rho.$
Here, $\alpha$ is the net round-trip field-amplitude
factor, including material gain or loss and radiative
attenuation, whereas $n_{\mathrm{ring}}^{\mathrm{eff}}$
is the real effective index governing the propagation
phase.
The fitted parameters are treated as constant over
the wavelength interval considered.

The fit gives $\tau=0.86142$ and
$n_{\mathrm{ring}}^{\mathrm{eff}}=1.76206$.
The COMSOL  can give
$n_{\mathrm{bus}}^{\mathrm{eff}}=1.92149$.
Figure~\ref{app2} compares the simulated transmission
coefficient with the fitted model, showing its
magnitude, imaginary part, and real part.
The model reproduces the principal features of
the complex transmission spectrum.

We repeat this procedure for different
coupling gaps and imaginary parts of the ring-core
refractive index.
The coupling gap primarily controls $\tau$ and
$\kappa$, whereas the imaginary part of the material
index primarily controls the round-trip field amplitude factor $\alpha$.
Within the coupling regime considered here, decreasing
the gap reduces $\tau$.
Increasing material loss decreases $\alpha$, while
increasing material gain increases it.

For the CPAL demonstrated in Figs. \ref{fig5} (a), (c), and (d), we list the corresponding material and geometry parameters:\\
  \begin{tabular}[c]{|l|l|}
    \hline
Lossy ring  & Gain ring   \\
    \hline 
 $d_{gap}=550[nm]$ &  $d_{gap}=550[nm]$ \\
   $n_{core}^{ring}=2.5+0.00511886i$ & $n_{core}^{ring}=2.5-0.00613281i$  \\
   $n_{core}^{bus}=2.5 $ & $n_{core}^{bus}=2.5 	$\\
   $n_{cladding}=1.5$ &  $n_{cladding}=1.5$\\
   $w=0.2\mu m$ & $w=0.2\mu m$\\
   $\rho=6.2 \mu m$ & $\rho=6.2 \mu m$\\
      \hline
\end{tabular}\\

For $T=0$ demonstrated in Figs. \ref{fig5} (b) and (d), we list the corresponding material parameters:\\
  \begin{tabular}[c]{|l|l|}
    \hline
Loss ring  & Gain ring   \\
    \hline 
 $d_{gap}=580[nm]$ &  $d_{gap}=580[nm]$ \\
   $n^{ring}=2.5+0.0029i$ & $n^{ring}=2.5-0.0038i$  \\
   $n_{core}^{bus}=2.5 $ & $n_{core}^{bus}=2.5 	$\\
   $n_{cladding}=1.5$ &  $n_{cladding}=1.5$\\
   $w=0.2\mu m$ & $w=0.2\mu m$\\
   $\rho=6.2 \mu m$ & $\rho=6.2 \mu m$\\
     \hline
\end{tabular}\\

 \end{document}